\documentclass[12pt]{article}   
\begin{document}

\noindent
{\scriptsize{22nd.August 2012}}

\noindent
{\bf\large  Three small systems showing probable 
room-temperature superconductivity}

\vspace{2em}
\noindent
{\bf D.M. Eagles$^*$}        

\vspace{1em}
\noindent
{\em 19 Holt Road, Harold Hill, Romford, Essex RM3 8PN,
England}

\vspace{2em}
\noindent
{\large ABSTRACT}

\vspace{1em}
\noindent
I shall discuss three small systems in which I think room-temperature
superconductivity has been observed, in the hope that experimentalists
will be persuaded to try to reproduce and extend results on some or all
of the three systems mentioned.  These are: 1. Narrow channels through
films of oxidised atactic polypropylene (OAPP) and other polymers.
2. Some multiwalled carbon nanotubes or mats of nanotubes.  3. Sandwich
structures based on CdF$_2$.  The main emphasis will be on polymer films.

\vspace{1em}
\noindent
Keywords: Room-temperature superconductivity, bipolarons, polymer films,
carbon nanotubes, CdF$_2$ sandwich structures. 

\noindent
PACS: 73.61.Ph, 74.20.Mn, 74.78.Bz, 71.38.Mx

\vspace{1em}
\noindent
{\em * E-mail address:} d.eagles@ic.ac.uk

\vspace{2em}

\noindent
{\bf 1. Introduction}

Small systems or systems with small subsystems are advantageous
for superconductivity in two ways: First because it is easier to get
Bose-Einstein condensation if there is separation in energy between the
lowest state of the system and excited states; secondly because it is
easier to get pairing at low carrier concentrations in low dimensions,
since, in the BCS regime there is an enhanced density of states at low
carrier concentrations in one and two dimensions compared with three,
and, in the BEC regime because bipolarons can form more easily in low
dimensions.  The separation in energy between the ground state and excited
states mentioned in the first point above is normally only a significant
effect if the density of electronic states as a function of energy is
such that condensation cannot occur in an infinite system, but in such
cases it can have a large effect.   For instance, for Bose condensation in
small arrays of filaments if a simple quadratic dispersion is assumed
for the bosons, then the condensation temperature at small lengths $L$
can increase approximately as $(1/L)$, as may be seen from Eqs. (21)
and (2) of \cite{Ea98}.  On the negative side it is not possible to
get infinite conductivity in quasi one-dimensional systems because
of phase slips, and only a small fraction of a Meissner effect is
possible if the transverse dimensions are of the order of or smaller
than the magnetic-field penetration depth.  Also on the negative side,
superconductivity in small systems will only have small-scale applications
such as switching devices or interconnections in electronics, unless small
systems can be combined weakly into larger ones without losing the advantages
of the original small sizes.

Two of the small systems I shall mention involve arrays of quasi
one-dimensional systems, although in one of the cases the existence of these
arrays is based on theory rather than experiment, and the third is a
quasi two-dimensional sandwich system.   The dates of the first claims of
room-temperature superconductivity in the three systems are separated by
about ten years, the first in 1988 and published in the open literature
in 1989, the second in 1999, and the third in 2009.

\noindent
{\bf  2. Two thin-film systems and one nanotube-based system}

In 2009, 2010 and 2012,  Bagraev and coworkers gave evidence
for superconductivity above room temperature in a CdF$_2$-based sandwich
system, involving two 3 nm thick boron doped layers sandwiching a 2
nm thick $p$-type layer of CdF$_2$, all on top of a 1 mm thick $n$-type
substrate \cite{Ba09,Ba10a,Ba10b,Ba12}.  Their work includes
evidence for zero resistance at 319 K, large diamagnetism starting
at somewhat higher temperatures, tunnelling evidence of an energy gap
of about 102 meV, and peaks in the specific heat at temperatures close
to those of the resistance drop.

After first suggestions of possible room-temperature superconductivity in
multiwalled carbon nanotubes by Tsebro and coworkers \cite{Ts99}, Zhao
and coworkers gave evidence from 2001 \cite{Zh01,Zh06,Zh08} from
their own results and published data of others that superconductivity
occurs for temperatures up to over 700 K in multiwalled nanotubes
or mats of single-walled or multiwalled nanotubes.  The evidence
is from resistance analysed by a slightly modified version of the
Langer-Ambegoakar-McCumber-Halperin (LAMH) theory of resistance due to
phase slips, from diamagnetic susceptibility much reduced from a full
Meissner effect due to magnetic field penetration depths being larger
in magnitude than tube diameters, tunnelling results showing large
energy gaps, and analysis of the temperature dependence of Raman data
for phonons of energy close to the energy gap.  In a review paper
\cite{Zh06}, Zhao gives reasons for the extraordinarily wide variation
of $T_c$ in different nanotubes or nanotube mats.  The main contibutions
to the variation are: 1. Different Luttinger-liquid parameters due to
different amounts of screening because of different surroundings of any
nanotube;  and 2. Variations in carrier concentrations.

There are five reasons for thinking that narrow channels through films
of oxidised atactic polypropylene (OAPP) are superconducting at room
temperature:

\noindent
1. Measurements with In microcontacts on top of films on a Cu substrate show
that at some points resistance is fairly low, and, within the spread of values,
is approximately independent of the film thickness \cite{Gr88,En89}.
There is also evidence \cite{En89} that superconductivity persists up
to at least 429 K in currents of 0.5 A.

\noindent
2. Estimates of the conductivity by direct \cite{Ar90} and indirect
\cite{De90} methods are several orders of magnitude larger than that
of Cu.  I have some difficulty in understanding how the  authors of
\cite{Ar90} obtained their direct results on conduction by four-probe
measurements, but I am more confident in the indirect estimates of
conductivity discussed in 4. below.

\noindent
3. The high conductivity is destroyed by non-thermal means by pulsed
currents of the order of 60 A through channels or groups of channels,
implying that the critical current density lies between 10$^8$ A
cm $^{-2}$ and about 5$\times 10^9 $ A cm $^{-2}$, depending on whether
the whole contact area or the estimated diameter of a single channel
is used to convert a critical current to a critical current density.
Probably the density is somewhere between these values, because
the method of measuring with pulses of gradually increasing magnitude  may
cause several channels to merge before the critical current is reached.

\noindent
4.  The thermal conductivity measured with contacts to the channels is
not significantly different from that with contacts to insulating regions
of the films \cite{Gr91a}, implying that the electronic contribution to
the thermal conductivity is very small, and that the the Wiedemann-Franz law
is violated by several orders of magnitude, as expected for a superconductor.

\noindent
In the pulsed current measurements mentioned above, with use of the
thermal conductivity results, it is estimated that,
for pulse lengths of the order of a microsecond, the heat front could
only propagate 0.3 $\mu$m into the channel, a much smaller distance than
the channel length of about 30 $\mu$m.  The heating of the channel must
be less than about 1000 K since otherwise the polymer would decompose
in 1-10 seconds at  a pulse repetition rate of 25 Hz used in some of
the pulsed measurements, and this leads to an indirect estimate of the
conductivity in terms of the pulse length, diameter of the channel, and
the critical current, which turns out to be several orders of magnitude
greater than that of Cu.

\noindent
5. There is large diamagnetism in some samples of OAPP at low fields  as
shown by magnetisation measurements \cite{En89,Ro00} and by occasional
spontaneous jumps of films to lower-field regions in inhomogeneous fields
\cite{Gr93}.  Both effects are thought to be related to superconducting
channels which join at their ends to form closed loops.  Only a small
diamagnetism can be expected due to the Meissner effect for the systems
being studied, both because there is never more than about 1\% of the film
occupied by conducting channels and because there is expected to be a
further reduction by a factor of the order of 100 if the magnetic-field
penetration depth is of the order of the channel radii.  However, I have
been able \cite{Ea02} to interpret details of the magnetisation versus
field curves obtained by Rogachev and Grigorov in 2000 \cite{Ro00} for
fields below a metamagnetic transition in terms of a theory of Shoenberg
\cite{Sh52} for induced currents in closed loops of type I superconductors
below the critical field.  This was assumed in \cite{Ea02} to apply
to type II superconductors with no pinning at fields below $H_{c1}$,
but I now think that it will also apply even with pinning for fields
below the field $H_R$ at which resistance arises.  The fits with a
model with two types of loops to adjusted data  after corrections for
an estimated superparamagnetic contribution from channels which do not
form closed loops deduced from results for magnetism on the next day
after large fields had been applied are shown in Fig. 1 of \cite{Ea02}.
Two results of interest which come out of the fitting and analysis are
that $H_R\approx$ 5260 Oe, and that the diameter of the cross section
of the channels is less than about 1.52 $\mu$m, compared with a
value of about 1 $\mu$m estimated by Grigorov and coworkers in 1990
\cite{De90}.  The initial susceptibility in the first film corresponds to
about 0.7 \% of a full Meissner effect, and I know of no mechanism not
involving superconductivity which can give such a large diamagnetism.
Estimates of $H_{c1}$ based on theory of Alexandrov \cite{Al93,Al04}
for bosonic superconductors combined with parameters obtained in the
latest version \cite{Ea11}\footnote{The bracketing in Eq. (1) of
this paper was incorrect.  Also the directions of the aligned dipoles
were stated incorrectly in lines 3 and 9 of section 4,  and, in subsection
14.14, last line, "greater than 1.52 $\mu$m" should be replaced by
"less than 1.52 $\mu$m".  An erratum has been submitted about Eq. (1).
The correct version of this equation may also be seen in v2 of the
preprint arXiv: 1106.0716} of my model for superconductivity in OAPP are
much smaller than the value of $H_R$ given above, and I now interpret
the value as that of the depinning field.

For the diamagnetism and for metamagnetic transitions which occur both in
OAPP and in  polydimethylsiloxane (PDMS) there are often time-dependent
effects over periods of days, and application of a field above a critical
value may disrupt this time dependence (see e.g. \cite{Gr96}).

The theory of Grigorov and coworkers \cite{Gr90a,Gr90b,Gr91b} for why
conducting channels exist in polar elastomers such as OAPP involves an
unusual type of polaron where rotatable dipolar groups become aligned
within a certain radius towards or away from an excess charge.  Also,
for a high ratio of static and high-frequency dielectric constants,
these polarons can coalesce to form linear strings called superpolarons,
and the superpolarons coalesce to form conducting channels.

The latest version of my model for the superconductivity \cite{Ea11}$^{1}$
is modified considerably from an early version \cite{Ea05}.  Both versions
involve Bose condensation in an array of nanofilaments.  Two differences
from the earlier version are: 1. An assumed Bogoliubov form for the
dispersion of the bosons, with an effective bosonic potential $\mu_B$
at $T_c$ which is considerably smaller than that at $T=0$; 2. Larger
numbers of nanofilaments in the smallest superconducting channels, based
on an estimate of the resistance for an N-S-N system with good contacts,
where the superconducting system consists of an array of nanofilaments
(cf. \cite{Fe04}), and comparison with resistance histograms mentioned
before.  Although there can be pairing in a single nanofilament, the usual
objections to long-range order in one dimension prevent the occurrence
of superconductivity.  Also, since the fractional reduction of $T_c$
due to resistance from phase slips in arrays depends on (the array cross
section)$^{-2/3}$ \cite{La67a} at least for BCS superconductivity,
small arrays cannot produce high-$T_c$'s.

My model makes a lot of use of a 1991 paper by Grigorov \cite{Gr91b}
for ferromagnetic superpolarons, and assumes that pairs form in the
superpolaron strings by attraction via either plasmons or high-frequency
phonons.  Most of the parameters are determined by minimising a weighted
sum of moduli of differences between calculated values and best-fit-values
of parameters occurring in equations determining theoretical or
experimental constraints, with weightings determined by how accurate I
think the constraints are.  For the parameters I find, the limiting $T_B$
for large $n_T$ is not much higher than room temperature, and so, as in an
earlier paper \cite{Ea05}, to explain the apparently considerably higher
$T_c$'s which occur when significant currents are applied, we have to
invoke decreases in transverse masses due to current-current interactions
decreasing the transverse lattice constant in the soft materials being
discussed.  Alternatively it is possible that a method I used to estimate
$\mu_B$ gives too high a value.  For a much smaller $\mu_B$ we would
find a relatively high value of $T_B$ for large-diameter channels.

There are some similarities and some differences from treatment of arrays
of nanowires by Bianconi and coworkers \cite{Pe96,Bi98}.  Differences
in their treatment include: 1. Coupling  weak enough for the carrier
concentrations studied for a BCS-like theory to apply, as opposed to the
BEC regime considered by me. 2. Parameters such that the second or third
quantised level with respect to transverse motion in a nanowire becomes
occupied, whereas in my model for OAPP only the ground state is occupied.
(For the model of Grigorov for nanofilaments in OAPP, the Fermi energy
before pairing is approximately proportional to the square of the
reciprocal of the nanofilament radius, and it is not possible to find a
situation in which the second quantised level becomes occupied, except
possibly for very thin films in which charge injection may increase the
carrier concentration).  

We make conjectures about the reasons for stronger pairing and higher
$T_c$'s in OAPP than in established high-temperature superconductors.
A first difference is the smaller high-frequency dielectric constant
(2.2 in polypropylene) which permits stronger coupling to plasmons (see
e.g. \cite{Bo05}), and also, if combined with a high static dielectric
constant, stronger coupling to polar optical phonons.  A second difference
is that there is not expected to be any significant increase in masses
due to correlation, permitting a higher condensation temperature in the
BEC regime for a given carrier concentration and strength of coupling
with bosons.  A third difference is that there are very high energy
phonons in polypropylene (up to 0.39 eV), which is helpful if pairing
is mediated by phonons.  Eight suggestions of properties useful for very
high-temperature superconductivity were made on pp 1945-6 of \cite{Ea05}.

The parameters of the model I used for OAPP indicate that this
system is on the BEC side of the BCS-BEC crossover, but not very
far from it.  Maximum $T_c$'s near the crossover are common to many
models. The crossover in superconductors was first studied in 1969
\cite{Ea69}\footnote {I should like to give a belated acknowledgement
to G.L. Sewell for a useful conversation in the mid 1960's, in which he
drew my attention to the fact that pairing does not necessarily give
rise to superconductivity.} after work of Labb\'e et al. in 1967 \cite{La67b}
had drawn attention to the fact that, for strong pairing and low carrier
concentrations, shifts in the Fermi level due to pairing have to be
considered.  In 1980 Leggett \cite{Le80} discussed the crossover in an
atomic system.  Studies of the crossover have proliferated since 2004 when
it was observed by several groups in atomic gases (see e.g. references
in \cite{Fa07}).

The first claim of a BEC superconductor was made by Ogg in 1946
\cite{Og46}.\footnote{A few years ago a large group from University
College, London, was trying to reproduce Ogg's results, but does not
appear to have published any results.}  Early indications of a
superconductor on the BEC side of the transition were obtained in 1986
\cite{Ta86,Ea86,Ea89} in one ceramic sample of SrTiO$_3$ with 3\% of
Ti replaced by Zr.  Some authors (see e.g. \cite{Ch05,Al07}) think that
at least the underdoped cuprates are BEC superconductors.  Most recent
evidence of a crossover has been found for one of the bands in multiband
superconductors due to the Fermi level being close to the bottom of a
second band in a quantum well \cite{In10}, or for small occupation of
one of the bands for some other reason \cite{Lu11}.

It may be of interest to speculate as to why the experimental results
on the systems mentioned have not been reproduced by other groups.
We suggest that this is a due to a combination of some or all of the
following factors: (a) Not many people being aware of the results;
(b) Those that are aware not believing the interpretations given; (c)
The commonly held belief that, if results are not confirmed quickly,
then they are not likely to be valid; and (d) Some prejudice against the
possibility of room-temperature superconductivity.  In the case of Bagraev
and coworkers it is probable that factor (a) is dominant.  Three of the
four papers of theirs that I mentioned appeared in a journal that is
probably not read widely in the superconductor community, and none of
the four papers mention superconductivity in their titles.  In the case of
Zhao and coworkers, since they include experimental results of others in
their analyses, factors (b) and (d) seem more likely.  A difficulty, felt
by me at one stage, is the claimed extraordinarily wide variation of $T_c$
between different nanotube-based systems, but Zhao addresses this problem
in his review article \cite{Zh06}.  Zhao has a good reputation for work
on the cuprates, but, despite he and coauthors writing preprints since
2001 on room-temperature superconductivity in nanotube-based systems,
only one journal paper by them has appeared on the subject up till now.
In the case of Grigorov and coworkers, I think it is probable that there
is a more even distribution between the four factors.

\noindent
{\bf 3. Conclusions}

To summarise, room-temperature superconductivity appears to have
been observed in several small systems.  I think it is important that
experimentalists should try to reproduce and extend the results.  For each
of OAPP and the CdF$_2$-based sandwich system, most of the results have
been obtained by only one group, although A.N. Ionov and coworkers have
done a lot of work since 1999 confirming that superconductivity occurs in
superconductor-polymer film-superconductor systems at temperatures below
the superconducting $T_c$ of the electrodes, and find that this is not due
to the proximity effect. (See e.g. \cite{Io10} and references therein
and in my 2011 paper).  A recent paper on PDMS not referred to in
\cite{Ea11} is \cite{Ha10}, where the authors search for conductivity
through films of PDMS of thicknesses 1 to 3 $\mu$m with various types of
electrodes.  They observed conduction with resistances of the order of 1
to a few $\Omega$ for some electrode configurations but not for others.
It appears that whether conduction is seen depends on whether or not the
electrode configuration is such that the electrodes exert some pressure
on the films.  This is consistent with the view of Grigorov and coworkers
(see e.g. \cite{Gr90a,Gr90b}) that pressure tends to align conducting
channels through the film.

Personally I think that, of the three systems I have discussed, if one
is prepared for some time-dependent properties, the materials problems
seem to be easiest for the polymers.  Some suggestions for possible new
experiments on OAPP films are made in subsection 14.15 of \cite{Ea11}.

\noindent {\bf Acknowledgement}

I should like to thank A.S. Alexandrov
for correspondence regarding $H_{c1}$ in bosonic superconductors.

\end{document}